\documentstyle[11pt,cospar,psfig]{article}

\newcommand{\etal}      {et al.\ }
\newcommand{\BoMo}      {Bockel\'ee-Morvan}

\def\lesssim{\mathrel{\hbox{\rlap{\hbox{\lower4pt\hbox{$\sim$}}}\hbox{$<$}}}}
\def\gtrsim{\mathrel{\hbox{\rlap{\hbox{\lower4pt\hbox{$\sim$}}}\hbox{$>$}}}}

\title{BIOMOLECULES IN THE INTERSTELLAR MEDIUM AND IN COMETS}

\author{S.B. Charnley\address{Space Science Division,
MS 245-3, NASA Ames Research Center, Moffett Field CA 94035-1000,
USA}, S.D. Rodgers$^1$, Y.-J. Kuan\address{National Taiwan Normal
University \& Academia Sinica Inst.\ of Astron.\ \& Astrophys., Taipei,
Taiwan}, and H.-C. Huang$^2$ }

\begin{document}

\maketitle

\begin{abstract}

We review recent studies of organic molecule formation in dense
molecular clouds and in comets.  We summarise the known organic
inventories of molecular clouds and recent comets, particularly
Hale-Bopp.  The principal chemical formation pathways  
involving gas phase reactions, as well as formation by catalytic reactions on 
grain surfaces or through dust fragmentation,  are identified
for both dense clouds and cometary comae.  The processes leading to
organic molecules with known biological function, carbon chains,
deuterium fractionation, HNC and S-bearing compounds are described.
Observational searches for new interstellar organics are outlined and
the connection between observed interstellar organics and those
detected in comets Hale-Bopp and Hyakutake are discussed.

\end{abstract}

\section*{INTRODUCTION}

The birth sites of solar-mass stars and their planetary systems are
dense interstellar clouds. Such clouds provide the building blocks, in
the form of interstellar molecules and dust, for the formation of
planets, comets, asteroids, and other macroscopic bodies in
protostellar disks (Ehrenfreund \& Charnley 2000; Irvine \etal 2000;
Lunine \etal 2000). These clouds are the site of an extremely active
and complex chemistry; to date over 100 molecules have been detected
in them. Many of these molecules are also observed in protoplanetary
disks and comets. A central question in the fields of astrochemistry
and astrobiology is how much of the material present in primitive
bodies, such as comets and asteroids, is pristine interstellar material?
In other words, to what extent does the chemical inventory of these
objects and the planets reflect chemical processing, firstly in the
collapsing cloud, and subsequently in the protoplanetary disk (e.g.\
van Dishoeck \& Blake 1998; Langer \etal 2000)?

Answering this question is of fundamental concern to two areas of
research. Firstly, chemical abundances, together with isotopic and
ortho-para ratios, are sensitive tracers of the molecule formation
conditions and so differences between the chemical inventory of the
interstellar medium (ISM) and the solar system can be used to
constrain models of low-mass star formation, many aspects of which are
not well understood. Secondly, the molecules in the early solar nebula
provided the seed from which life eventually formed. It is thought
that a great deal of organic material was deposited on the young Earth
by cometary and asteroid impacts (Chyba \etal 1990). Many of the
organic molecules detected in dense interstellar clouds have important
functions in terrestrial biochemistry.  The presence of
extraterrestrial amino acids in meteorites (e.g.\ Cronin \& Pizzarello
1983) shows that extremely complex molecules are able to form in
space. In order to assess the implications for life elsewhere in the
universe, we need to understand if such complexity is a general
feature of interstellar and/or protoplanetary disk chemistry, or if
these molecules arose due to specific conditions in our solar
system. We can obtain clues to the above question by comparing
observations of comets -- thought to be the most pristine objects in
our solar system -- with those of protoplanetary disks, protostellar
cores, and interstellar clouds (Ehrenfreund \etal 1997; Ehrenfreund
2000; Ehrenfreund \& Charnley 2000; \BoMo\ \etal 2000; Irvine \etal
2000).

\begin{table}[t!]
\caption{ Representative compositions of the gas in a cold molecular
cloud (L134N), protostellar ices (NGC7538:IRS9), protostellar hot core
gas (Sgr B2(N)), and in a cometary coma (Hale-Bopp).}
\begin{center}\scriptsize
\begin{tabular}{lllll}
\noalign{\smallskip}
\hline \hline
\noalign{\smallskip}
Molecule& L134N & NGC7538:IRS9 & Sgr B2(N) & Hale-Bopp \\
\noalign{\smallskip}
\hline
\noalign{\smallskip}
H$_2$O &$<3$  & 100   & $>$ 100  & 100\\
CO     &1000 & 16  &  1000 &  20 \\
CO$_2$ & - & 20 & -  & 6-20\\
H$_2$CO &0.25 & 5 & $>0.005$  & 1\\
CH$_3$OH &0.04 & 5  & 2 & 2\\
NH$_3$ & 2.5 &13 & - & 0.7-1.8\\
CH$_4$ &- & 2 & - & 0.6\\
C$_2$H$_2$ & - & $<$10 & - & 0.1\\
C$_2$H$_6$ & -&  $<$ 0.4 & - & 0.3\\
HCOOH     &0.004 &  3 & $>0.003$ & 0.06\\
CH$_2$CO& $<$0.009& - &0.002& $<0.03$\\
CH$_3$CHO& 0.008 & - &0.002 & 0.02\\
$c$-C$_2$H$_4$O &-& - &0.001 &-\\
CH$_3$CH$_2$OH&- &$<$1.2 &0.01 & $<0.05$\\
CH$_3$OCH$_3$&-& - &0.03 & $<0.45$\\
HCOOCH$_3$ & $<$0.02 & - &0.02 & 0.06\\
CH$_3$COOH & - & - & 0.0008& - \\
CH$_2$OHCHO & - & - & 0.003 & - \\
OCN$^-$ & - & 1 & - & -\\
HNCO & - & - & 0.006 & 0.06-0.1\\
NH$_2$CHO  & $<$0.001 & - & 0.002 & 0.01\\
HCN   & 0.05 &- & $>0.05$  & 0.25\\
HNC &0.08 &-& $>0.001$  &0.04\\
CH$_3$CN &$<$0.01  & - & 0.3 & 0.02\\
CH$_3$NC & - & - & 0.015 & - \\
HC$_3$N  &0.002  & - &0.05 & 0.02\\
C$_2$H$_3$CN & - & - & 0.6 & - \\
C$_2$H$_5$CN & - & - & 0.006 & - \\
H$_2$S  & 0.01 & - & - & 1.5\\
OCS & 0.02 &0.05 & $>$0.02 & 0.5\\
H$_2$CS & 0.008 & - &0.2 & 0.02 \\
SO& 0.25 & - & 0.2 & 0.2-0.8\\
SO$_2$ & 0.005 & - & 0.3 & 0.1\\
\noalign{\smallskip} \hline
\end{tabular}
\end{center}
{\scriptsize NOTES: Abundances for interstellar ices and cometary
volatiles are normalized to H$_2$O. Gaseous abundances for Sgr B2(N)
and L134N are normalized to CO\@. Original references for the
abundances in L134N and NGC7538 appear in Charnley \etal
(2001b). Values for Sgr B2(N) and comet Hale-Bopp are taken from
Nummelin \etal (2000) and Crovisier \& \BoMo\ (1999) respectively.}
\end{table}

\section*{THE INTERSTELLAR-COMET CONNECTION}

Table~1 compares the abundances of a number of molecules in a dark
cloud, ices toward a protostar, a star-forming `hot molecular core'
where ices have recently evaporated, and a cometary coma. Whilst the
analysis is complicated by factors such as the probable depletion of
CO and CO$_2$ in cometary ices (Yabushita 1995), and the fact that
some of the molecules in hot cores and comets are likely to be
daughter species formed via post-evaporation chemistry in the warm
gas, it is clear that there are many fundamental similarities between
the different regions. However, there are also a number of differences
between cometary and interstellar abundances, which suggests that
whilst solar system objects may consist of some purely interstellar
material, other constituents have undergone significant processing. 
It is also apparent that there are substantial differences between
molecular abundances in interstellar dark clouds and warm protostellar
regions.  This can be explained in terms of surface chemistry
occurring as atoms and molecules freeze out onto dust grains in the
cold cloud, followed by a complex gas-phase hot core chemistry which
is initiated when the grain mantles are evaporated by the nearby
protostar. A similar chemistry is thought to occur as material falls
onto the disks surrounding low-mass stars, and the products will then
refreeze and be incorporated into cometesimals and planetesimals
(Lunine \etal 2000). 
Therefore, it is clear that in order to address the above question in
a quantitative manner, we require a detailed understanding of the
relevant chemical reactions and physical processing that interstellar
material experiences as it suffers the violent passage from a cold
dense cloud into a protoplanetary system.

\section*{ORGANIC ASTROCHEMISTRY: OBSERVATIONS AND THEORY}

\subsection*{Interstellar Medium}

\leftline{\underline{Cold Dark Clouds}}

The two most intensively studied dark clouds are TMC-1 and L134N (see Table 1 of
Ehrenfreund \& Charnley 2000). Because of the low temperatures in
these clouds, the chemistry is driven by exothermic ion-molecule
reactions, initiated by cosmic ray ionization. In general, chemical
models are able to give reasonable agreement with the observed
abundances, although there remain problems with sulphuretted molecules
and carbon-chain species (e.g.\ Millar \etal 1997). It is also
possible that grain surface chemistry may   contribute to the
formation of molecules seen in dark clouds. Although these clouds are
cold, a number of possible non-thermal desorption mechanisms have been
proposed to return material from grains to the gas phase (e.g.\
Markwick \etal 2000); these proposals are motivated by the fact that
in dense clouds one would expect heavy elements to be almost totally
depleted from the gas on timescales much less than the estimated ages of the
clouds (cf.\ Charnley \etal 2001a). 
The largest molecules observed in dark clouds are linear carbon chain
molecules, such as C$_8$H and HC$_{11}$N (e.g.\ Bell \etal 1997,
1999). However, the decline in abundance with increasing chain
length suggests that only a small fraction of interstellar carbon is
locked up in longer chains. Nevertheless, it is likely that a
significant fraction of the carbon is in the form of large aromatic
molecules (see below). 
After atoms and molecules accrete onto grains to form ice mantles, it is
likely that the resulting ices are subsequently processed by UV
photons and/or energetic cosmic ray impacts. Laboratory experiments involving
irradiation and radiolysis of interstellar ice analogs have
demonstrated that relatively simple ices can be processed into much
more complex molecules such as ethanol, formamide, carbonic acid,
lactic acid, urea, and even larger molecules (Agarwal \etal 1985;
Bernstein \etal 1995; Moore \& Hudson 1998). Greenberg (1998) has
proposed that over the lifetime of a molecular cloud a substantial
fraction of the ice is converted into complex organic material, and that
when planetary systems form this material is eventually incorporated
into comets.

\leftline{\underline{Hot Molecular Cores}\rule{0cm}{8mm}}

The first hot molecular cores to be discovered were the Hot Core and
Compact Ridge sources in Orion-KL, which are characterised by large
abundances of small saturated molecules (e.g.\ NH$_3$, H$_2$O,
CH$_3$OH) in warm ($T\geq100$\,K), dense ($n\sim10^7$~cm$^{-3}$) gas,
with anomalously high molecular D/H ratios for such high temperatures
(e.g.\ Blake \etal 1987). Many other hot cores have since been
identified and studied, and many molecules have been detected in them
(see Table~1). The relation of many hot cores with ultracompact
H\,{\sc ii} regions confirms that they are associated with the
earliest stages of high-mass star formation (Kurtz \etal 2000), and
their composition is thought to result from the evaporation of
interstellar ices by the newly-formed protostars (Brown \etal 1988).

In addition to water, ammonia and methanol, hot cores also contain
lesser abundances of larger saturated species such as ethanol and
ethyl cyanide, which are not detected in dark clouds.  On the other
hand, the long unsaturated chain species which are observed in dark
clouds, such as the larger of the cyanopolyynes, are absent from hot
cores.  This is generally taken to be evidence for a limited surface
chemistry occurring on interstellar grains. The prevalence of hydrogen
in the ISM, and the much greater mobility of H atoms on the grain
surface, ensures that hydrogenation is the principal reaction for
surface molecules (Tielens \& Hagen 1982); saturated alkanes and
alcohols are the end result of H addition to the linear chain
molecules observed in dark clouds (e.g.\ Charnley 1997a). After the
evaporation of the initial ices, a rich gas phase chemistry can occur,
and large abundances of new molecules can be formed (Charnley \etal
1992, 1995).  Although broadly similar, the detailed chemical
composition of the gas is seen to vary between different hot
cores. For example, the Orion Hot Core appears to be enriched in
N-bearing species, whereas the Compact Ridge has enhanced levels of
O-bearing species (Blake \etal 1987); similar spatial differentiation
is seen in a number of other sources (e.g.\ Nummelin \etal 2000).
Charnley \etal (1992) showed that the N/O differentiation can be
explained by differing ammonia/methanol ratios in the original ices,
since many of the large O-bearing species are formed in the hot gas
via reactions involving protonated methanol, and the presence of
ammonia suppresses the CH$_3$OH$_2^+$ abundance since NH$_3$ acts as a
sink for protons. Caselli \etal (1993) proposed a mechanism to account
for the initial differentiation in the ices, but recent observations
suggest that NH$_3$ and CH$_3$OH do in fact coexist in interstellar
ices (Gibb \etal 2000).  Rodgers \& Charnley (2001a) showed that the
N/O differentiation may actually be an age/temperature effect, since
`cool' ($T\approx 100$\,K) cores will eventually lose their NH$_3$
after $\sim10^5$ yrs, whereas at higher temperatures endothermic
reactions become rapid enough to ensure that the NH$_3$ abundance
never drops substantially, and hence large abundances of O-rich
organics never evolve. Similar age/temperature effects may also be
apparent in the sulphur chemistry, since the presumed parent molecule,
H$_2$S, is processed into HS, S, S$_2$, SO, SO$_2$, H$_2$CS, and CS,
the peak abundances of which depend on the core temperature (which
controls the O\,{\sc i} abundance) and age. Therefore, observed
abundance differences in sulphuretted hot core molecules may be used
to constrain the age of the cores, and place limits on the time-scale
for this phase of the star formation process (Charnley 1997b)

Finally, an alternative explanation for the chemical variety amongst
hot cores lies in the nature of the source which drives the
mantle evaporation and the gas phase chemistry. All the models
discussed above assume temporally constant temperatures and
densities. This will be (roughly) appropriate for cores which are
heated radiatively, but it is likely that many hot cores have
experienced the passage of shock waves caused by the impact of
protostellar outflows into the surrounding natal cloud material. In
this case the postshock temperature can reach values higher than
1000\,K (e.g.\ Draine \& McKee 1993), this permits many reactions
which are endothermic, or have activation energy barriers, to occur. For
example, Charnley \& Kaufman (2000) showed that the low abundances of
CO$_2$ observed in hot cores can be explained if these regions are
heated by shocks. Shock waves can also erode the refractory grain
cores and the release of silicon, observed to be highly depleted in
cold clouds, is now understood to account for the presence of SiO in
star-forming regions (Schilke \etal 1997). 
These considerations   ensure that a thorough understanding of
hot core chemistry requires detailed chemical modelling, since only when all
possible gas phase production routes have been investigated and ruled
out, is it possible to infer that a particular molecule is a parent in
the original ice. Through this type of modelling, we can gradually
build  up a secure inventory of interstellar ices (cf.\
Ehrenfreund \& Charnley 2000). Comparison of inferred ice composition
with that observed by the {\it ISO} satellite will allow us to
determine the processing that the ices undergo as they are irradiated
and evaporated by nearby protostars.

\subsection*{Comets}

The modern view of comets began with the work of Whipple (1950), who
proposed the `icy conglomerate', or `dirty snowball' model of comet
nuclei. Over the years this model has been refined, particularly after
the discovery in comet Halley of large, so-called `CHON', organic
particles (Kissel \etal 1986), and comets are now known to consist of
roughly equal parts by mass of ice (mainly water, with $\sim10\%$ CO
and CO$_2$), inorganic refractories (mainly silicates), and organic
refractories (CHON particles and other carbonaceous material)
(Greenberg 1998). The ices also contain trace amounts of other
molecules at the $\sim1\%$ level or less (Table~1) -- although only a
minor constituent of the comet as a whole, these species are very
important, since they can be used to trace the conditions under which
comets were formed.  The recent detection of Argon in comet Hale-Bopp
(Stern \etal 2000) is proof that this comet has undergone very little
thermal processing since its formation, as Ar sublimates at
$T\approx35$\,K\@.

Based on their orbital characteristics it is possible to divide comets
into two distinct populations: long- and short-period comets. A more
strict classification is based on the Tisserand invariant, T\@; comets
with $\rm T > 2$ (short-period comets) are thought to originate from
the Kuiper belt (located at a heliocentric distance
$30<r_h\lesssim 50$~AU), whereas long-period comets come from the Oort
cloud ($r_h \sim 5\times10^4$~AU).  The current consensus is that
Kuiper belt comets most probably formed near their current location,
whereas Oort cloud comets were formed somewhere in the giant planet
region of the protosolar nebula ($5\lesssim r_h \lesssim 30$~AU),
before being ejected to their present position via gravitational
interaction with these planets (Weissman 1999). If this picture is
correct, one would expect to see chemical differences between the two
populations. This indeed seems to be the case for carbon chains;
A'Hearn \etal (1995) showed that short-period comets are much more
likely to be depleted in C$_2$ and C$_3$ than long-period comets. More
recently, Mumma \etal (2000) observed ethane in the short period comet
Giacobini-Zinner, and showed that it too is depleted relative to the
long-period comet Hale-Bopp.

Our detailed knowledge of the molecular inventory of comets has been
revolutionised in the past few years by the apparitions of the bright
comets Hyakutake and Hale-Bopp; many of the molecules listed in
Table~1 were observed for the first time in Hale-Bopp (\BoMo\ \etal
2000). The use of interferometers allowed the first detailed maps to
be made of the distributions of molecules in the coma (e.g.\ Blake
\etal 1999; Kuan \etal 2001a). Mapping of ions such as HCO$^+$ also
allowed the structure of the cometary ionosphere to be determined
(e.g.\ Lovell \etal 1997), and showed that ion-molecule chemistry may
be important in the coma since, for example, HCO$^+$ is formed from
proton transfer to CO\@.  The results of this mapping showed that, in
addition to a nuclear source, some molecules also appeared to have an
extended source in the coma. Although this was previously known to be
the case for formaldehyde in comet Halley (Meier \etal 1993), it is
now known that CO, OCS, HCN, and HNC also had extended sources in
Hale-Bopp.  Various theories exist to explain these distributions; it
may be that these molecules are formed via photodissociation of some
unknown parent, or they may be sublimating directly from icy grains in
the coma. Alternatively, as in hot cores, these species and many
others could possibly have been produced via post-sublimation
gas-phase chemistry occurring in the coma.

In order to quantitatively assess the potential for chemical reactions
to alter the composition of the coma, we have developed a model of the
chemistry and physics in the coma (Rodgers \& Charnley 2001d).
Despite the differences in the physical parameters and the molecular
lifetimes, the chemistry taking place in the coma is analogous in many
ways to that which occurs in hot cores.  However, because the physical
conditions change throughout the coma, it is necessary to calculate
the hydrodynamics in tandem with the chemistry and this ensures that
the development of realistic coma models is a far more demanding task.
Briefly, the model simultaneously solves the conservation equations
for abundances, mass, momentum and energy appropriate for a steady
spherical outflow.  Endoergic chemical reactions driven by
suprathermal H atoms are included, along with all the important
microphysical processes that determine the heating and cooling in the
coma.  As described below, we have used this model to investigate
possible production routes for HNC and complex organics, as well as
the potential of gas-phase chemistry to alter D/H ratios in the coma
(Rodgers \& Charnley 1998, 2001b,c,d).

\section*{COMPLEX ORGANICS}

\subsection*{Interstellar Medium}

It has become clear in recent years that the ISM contains a large
population of polycyclic aromatic hydrocarbon (PAH) molecules (e.g.\
L\'eger \& Puget 1984; Allamandola \etal 1985), with a typical size of
$\sim$20--100 C atoms. PAHs are thought to account for $\sim$10\% of
the interstellar carbon budget, and are most likely formed in the
outflows of evolved stars (Frenklach \& Feigelson 1989). However, it
has not yet proved possible to definitively identify {\it specific}
PAH molecules.  The largest positively-identified molecule thus far
detected in the ISM is HC$_{11}$N (Bell \etal 1997). Yet, despite
containing 13 atoms, structurally this is a simple linear chain
molecule. In terms of structure the most complex interstellar
molecules are seen in hot cores, where molecules such as ethanol,
dimethyl ether, ethylene oxide, methyl formate, acetic acid, and
glycolaldehyde have been detected. The latter three molecules are
isomers of each other, as are ethanol and dimethyl ether. It is clear
that any comprehensive theory of interstellar chemistry should be able
to explain the prevalence of some isomers over others. In the previous
section we discussed how the molecular abundances in hot cores are
determined by two different chemical processes; surface chemistry in
the cold gas prior to the formation of the hot core, and hot
post-evaporation gas phase chemistry. Here we briefly describe a
unified scheme to explain the observed molecular complexity in these
regions (cf.\ Charnley 1997a, 2001; Charnley \etal 2001b).

Unconstrained grain surface reactions can potentially produce many
molecules (e.g.\ Caselli \etal 1993), however, such models cannot
account for the specific abundances seen in hot cores. We therefore
limit the surface chemistry to consist of atomic addition reactions
(one atom at a time), with the further constraint of {\it radical
stability} imposed on the intermediate organic radicals. This results
in surface production of alcohols and amines, although the existence
of energy barriers for H addition to double and triple bonds may mean
not all molecules are fully reduced, resulting in some aldehydes,
ketones, and nitriles. This scheme naturally accounts for the
methanol, formaldehyde, ethanol, ketene, acetaldehyde, vinyl cyanide,
and ethyl cyanide seen in hot cores (Charnley 1997a).  It can also
explain the existence of the ring molecule ethylene oxide
(c-C$_2$H$_4$O) via O atom addition to the C$_2$H$_3$ radical, which
is formed when H atoms react with acetylene.  If instead of an O atom
an N atom is added to $\rm C_2H_3 $, either methyl cyanide or
vinylimine can result; however a ring structure is also possible and
this results in the azirines 1-H azirine ( H($c$--C$=$N--C)H$_2$ ) and
2-H azirine ( H($c$-C$=$C--NH)H ).  Complete reduction via addition of
two further H atoms will result in aziridine ( H$_2$(C--NH--C)H$_2$ ).
We therefore searched for these molecules in several hot cores with
the NRAO 12m telescope (Kuan \etal 2001b; Charnley \etal 2001b).
Figures.\ 1 and 2 show tentative detections of the azaheterocycles 2-H
azirine and aziridine in Orion-KL, Sgr B2(N) and W51.

\begin{figure}
\psfig{file=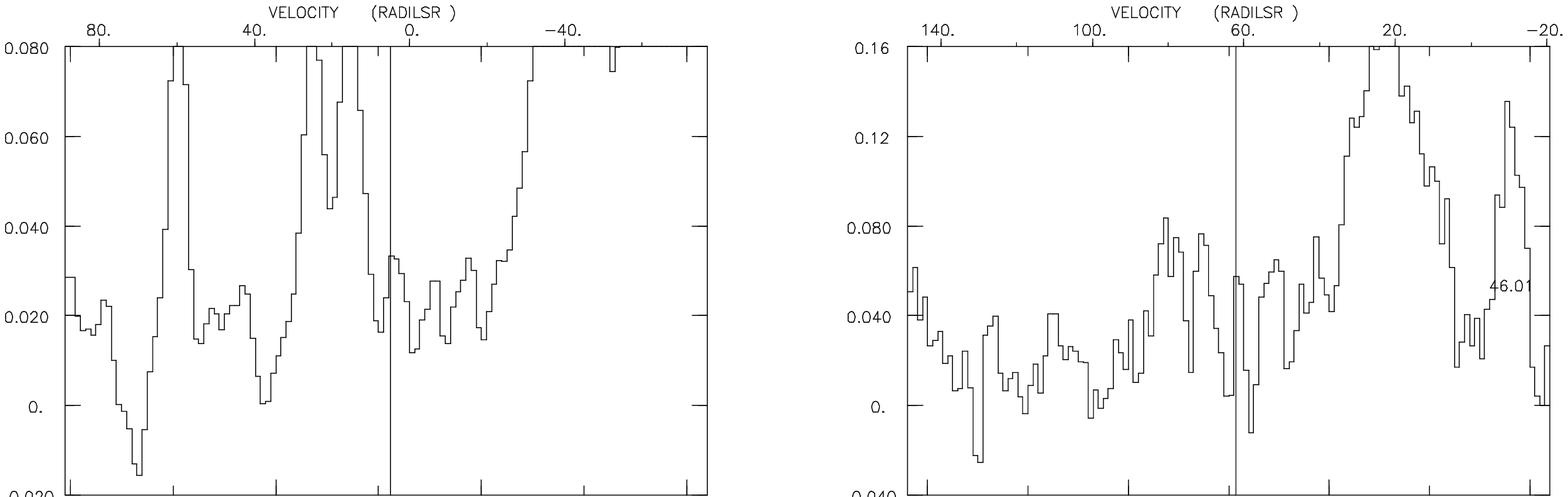,width=500pt}
\caption{2H-azirine 6$_{25}$--5$_{14}$ transition at 226.2 GHz. (a)
Orion-KL\@. (b) Sgr B2(N).}
\end{figure}

\begin{figure}
\psfig{file=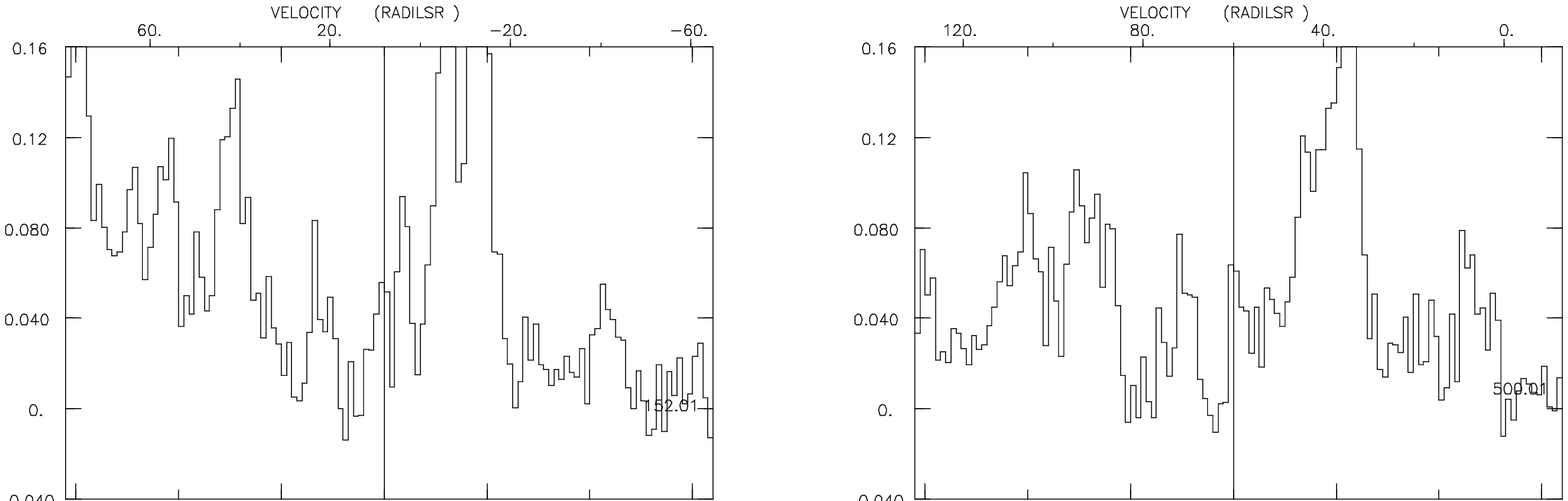,width=500pt}
\caption{Aziridine 6$_{15}$--5$_{25}$ and 6$_{25}$--5$_{15}$
transitions (blended) at 236.0 GHz. (a) Orion-KL\@. (b) W51 e1/e2.}
\end{figure}

Whilst surface chemistry can account for the molecules discussed
above, there are other molecules observed in hot cores that cannot be
explained by this mechanism, such as dimethyl ether and methyl formate.
Such molecules must therefore be formed in the warm gas; a plausible
mechanism is alkyl cation transfer reactions, whereby a protonated
alcohol transfers an alkyl group to a neutral base i.e.
\[
{\rm ROH_2^+ ~+~ R'X ~\longrightarrow~ R'XR ~+~ H_2O}
\]
Many of these reactions have been measured in the laboratory and found
to be rapid (e.g.\ Mautner \& Karpas 1986), and methanol and ethanol
are known to be abundant in hot cores. Thus, dimethyl ether can be
formed via self-methylation of methanol, and methyl formate may arise
from reactions of CH$_3$OH$_2^+$ with H$_2$CO or HCOOH (Charnley \etal
1995, 2001b). Transfer of methyl groups to other common hot core species can
produce many other large molecules, e.g.\ CH$_3$NCO, (CH$_3$)$_2$CO,
NH$_2$COCH$_3$, CH$_3$OC$_2$H$_5$,   isocyanides and large amides
(Ehrenfreund \& Charnley 2000; Charnley \& Rodgers, in preparation).
Another predicted product is diethyl ether, which was tentatively
detected by Kuan \etal (1999).

Finally, it is possible that the type of reactions we have discussed
may lead to glycine, the simplest amino acid. The surface chemistry
scheme we described can lead to aminomethanol, via hydrogenation of
HNCO\@. Alternatively, aminomethanol is thought to be formed when ices
containing H$_2$O, NH$_3$, and H$_2$CO are heated (Woon 1999). In
either case, one would expect significant abundances of aminomethanol
to be evaporated into hot cores, and subsequent protonation followed
by reaction with HCOOH may form glycine (Charnley 1997a).  Despite many
searches over the years, glycine has yet to be detected in the ISM,
but recent tentative identification of several lines suggest that it
may soon be conclusively identified (Charnley \etal 2001b).

\subsection*{Comets}

The brightness of comet Hale-Bopp means that it is the only comet in
which many large organic molecules have been positively detected. The
largest molecule observed in Hale-Bopp is methyl formate (HCOOCH$_3$)
(\BoMo\ \etal 2000), and Kuan \etal (2001a) reported a tentative
detection of its structural isomer acetic acid (CH$_3$COOH). As
discussed above, these molecules are also seen in hot cores where they
are thought to be daughter species.  Thus, it is possible that the
HCOOCH$_3$ (and CH$_3$COOH) seen in Hale-Bopp were actually created in
the coma. In order to investigate this possibility, we used our comet
model to follow the organic chemistry which occurs in the coma
following the sublimation of methanol rich ices from the nucleus
(Rodgers \& Charnley 2001b). We find that chemical reactions {\it
cannot} form sufficient quantities of HCOOCH$_3$ and other organic
molecules detected in Hale-Bopp: HCOOH, HC$_3$N and CH$_3$CN\@.
Therefore, these species must be present in the nuclear ice. Hence it
appears that cometary ices are more processed than their interstellar
counterparts, which is consistent with primordial interstellar ices
being evaporated, perhaps within the protosolar nebula, undergoing a
hot core-like phase of chemical processing, then refreezing onto
pre-cometary grains.  However, we cannot be absolutely certain that
these large organics are daughter molecules in hot cores, and they may
also be present in interstellar ices. One way to determine the origin
of these molecules may be to use D/H and $^{13}$C/$^{12}$C ratios
(Rodgers \& Millar 1996; Rodgers \& Charnley 2001e; Charnley \etal
2001c).

Photo-fragmentation of cometary CHON particles, first detected by the
{\it Giotto} probe in the coma of comet Halley, has been proposed to
account for the extended sources seen for many molecules and radicals
in the coma, in particular CO, C$_2$, H$_2$CO, CN and NH$_2$\@. 
Specifically,   to account for the coma sources of CO and
H$_2$CO,  and also the regular peaks in the mass spectra of CHON
particles  seen by {\it Giotto}, it has been proposed that a
substantial fraction of the CHON material consists of the formaldehyde
polymer, polyoxymethylene (POM, [--CH$_2$--O--]$_n$) (e.g.\ Huebner 1987).  
Other potential candidates  include hexamethylenetetramine
(HMT, C$_6$H$_{12}$N$_4$; Bernstein \etal 1995),
polyaminocynaomethylene (PACM, [--(NH$_2$)C(CN)--]$_n$; Rettig \etal
1992), and PAHs (specifically phenanthrene, C$_{14}$H$_{10}$; Moreels
\etal 1994). It is likely that a mixture of many types of organic
particles are actually present in comets, but we may be able to
constrain the amounts of the different types if we can model the
predicted production rates for daughter molecules, and compare them
with the observations.

\section*{THE HNC/HCN RATIO}

\subsection*{Interstellar Medium}

The HNC/HCN ratio varies a great deal in the ISM, with an inverse
temperature dependence: in cold dark clouds the ratio is typically
$\gtrsim 1$, whereas in hot cores it is low. This was previously
thought to be caused by three factors; i) low temperature ion-molecule
reactions preferentially forming HNC via the H$_2$NC$^+$ ion, ii)
proton transfer reactions cycling between the two isomers via the
HCNH$^+$ ion, and iii) high temperature isomerization reactions
transforming HNC into HCN\@. However, recent quantum chemical
calculations have showed that assumptions (i) and (iii) are incorrect
(e.g.\ Talbi \etal 1996).

In hot cores, HCN and HNC are thought to be daughter species, since
HCN is produced efficiently from reactions of NH$_3$ and C$^+$; at
high temperatures ($>250$K) the reaction of CN with H$_2$ ensures that the net
destruction rate of HCN is extremely small (Rodgers \& Charnley
2001a). Proton transfer to HCN to form HCNH$^+$ followed by
recombination or proton transfer to ammonia forms HNC, and 
models predict HNC/HCN ratios of $\sim 0.007$--0.05 for a 300\,K core,
depending on the age and the initial NH$_3$ abundance. This is in
reasonable agreement with the observed value in these regions, but we
are still unable to explain the excess of HNC in cold clouds.

\subsection*{Comets}

HCN has been seen in many comets, but HNC was detected for the first
time in comet Hyakutake, where it was present with an abundance
relative to HCN of 6\% (Irvine \etal 1996). HNC was subsequently
detected in comets Hale-Bopp and Lee, and the HNC/HCN ratio in Hale-Bopp
showed a strong increase as the comet approached perihelion, from
$<0.02$ at 2.5~AU to $\sim0.16$ at 1~AU (Irvine \etal 1998). The
Hyakutake results were originally interpreted as proof that cometary ices
contain unprocessed material from the ISM, but the observations of
Hale-Bopp proved conclusively that HNC must be a daughter species,
and that its production is related to the solar photon flux and/or
coma temperature.

We have used our comet model to investigate possible chemical
production routes to HNC in the coma (Rodgers \& Charnley 1998,
2001c). We find that ion-molecule chemistry is unable to synthesise
the observed quantities of HNC, but that endothermic isomerisation
reactions of HCN, driven by suprathermal H atoms produced in the
photodissociation of parent molecules, may be efficient in large,
active comets such as Hale-Bopp. However, in comets Hyakutake and Lee
we demonstrated that HNC must be produced via the photodestruction of
some unknown parent.  Possible molecular candidates are HNCO and
CH$_2$NH, but the former is thought to be photodissociated into NH $+$
CO and/or H $+$ OCN, whereas the latter is not observed in
comets. Hence, we consider it most likely that HNC is coming from CHON
particles in the coma, and conclude that the most likely candidate is
PACM (cf.\ Rettig \etal 1992). This source of HNC may also account for
the extended sources of HCN and CN in the coma.

\section*{CARBON CHAINS}

\subsection*{Interstellar Medium}

Carbon chain molecules appear to be almost equally abundant in both
the diffuse ISM and dense clouds (Allamandola \etal 1999), and it has
been proposed that C-chain anions may be responsible for some of the
diffuse interstellar bands (Tulej \etal 1998). The overall abundance
of long C-chains ($\gtrsim 10$ C atoms) in diffuse clouds is
$\sim10^{-10}$ with respect to hydrogen, and they are thought to
account for a few parts per million of the total cosmic carbon. In
comparison, $\sim10\%$ of the total cosmic carbon is in PAHs (see
above). Carbon chains are thought to be formed via neutral-neutral
C-insertion reactions involving CN and C$_2$H radicals (Fukuzawa \etal
1998).

The overall abundances of chain molecules are derived from infrared
observations of C$=$C vibrational transitions; in order to identify
specific molecules one must look at rotational transitions at
millimeter wavelengths. The largest molecule identified to date is
HC$_{11}$N which was seen in the dark cloud TMC-1 (Bell \etal
1997). The abundances decrease with increasing chain length, and this
detection is at the limit of current telescope capabilities, making the
detection of longer chains difficult in the near-term.  However, in recent
years the spectra of a large number of chain molecules have been
accurately characterised in the laboratory (e.g.\ McCarthy \etal
2000), and so future observations may be able to distinguish the precise
molecular nature of the smaller chains ($\leq10$ C atoms).

\subsection*{Comets}

Although there is no direct evidence for carbon chains in comets, the
presence of the C$_2$ and C$_3$ radicals suggests that they are fairly
abundant. The longest chain molecule positively detected in a comet is HC$_3$N,
which is   a parent species in Hale-Bopp (Rodgers \& Charnley
2001b). This suggests that longer cyanopolyynes may also be present
in the nucleus, albeit in lower abundances.  
It is interesting that
C$_2$ and C$_3$ are observed to be depleted in some Kuiper belt comets
(A'Hearn \etal 1995), implying that C-chain molecules are more
abundant in Oort cloud comets. This can be explained if these chains
were formed more efficiently in the warmer, giant planet region of the
protosolar nebula than the Kuiper Belt region. Alternatively, chain
molecules may form readily in both regions, but undergo further
processing in the outer region, perhaps involving saturation or the formation
of ring molecules and fullerenes. 
We previously discussed the probable contribution of polymeric
material to the CHON particles in comets (e.g.\ Huebner 1987; Rettig
\etal 1992). However, this material cannot be unequivocally identified
via spectroscopy of the coma, and we must wait until the {\it
Stardust} mission returns a sample of comet dust the the Earth for
laboratory analysis before we can properly understand the nature of
the CHON grains.

\section*{DEUTERIUM FRACTIONATION}

\subsection*{Interstellar Medium}

The overall D/H ratio in our galaxy is $\approx1.6\times10^{-5}$, but
the D/H ratios in some molecules can be up to 10$^4$ times larger,
since at low temperatures small zero-point energy differences result
in the fractionation of ions such as H$_2$D$^+$, C$_2$HD$^+$, and
CH$_2$D$^+$. Ion-molecule chemistry spreads the D-enrichment to a
variety of molecules, and the observed D/H ratios can be used to
constrain the gas temperature, the electron fraction, and the
depletion of heavy elements (e.g.\ Millar \etal 2000). In addition,
spatial variations in D/H ratios may allow the propagation of Alfv\'en
waves to be traced in molecular clouds (Charnley 1998). 
The D/H ratios observed in interstellar molecules can also be used to
probe the chemical mechanisms which form those molecules. Rodgers \&
Millar (1996) showed that D/H ratios in hot cores trace the values in
the precursor ice, and Charnley \etal (1997) used this fact to show
that the hot core abundances of the two isotopomers of deuterated
methanol -- CH$_2$DOH and CH$_3$OD -- should always have a ratio of
3:1 if methanol is formed on grains via hydrogenation of CO\@. Rodgers
\& Charnley (2001e) considered NH$_3$ production in dense cores, and
showed that although the total fractionation of ammonia may vary from
source to source, for a specific chemical production mechanism the
{\it ratio} of the multiply deuterated isotopomers should be constant,
and may be used to distinguish between gas phase and surface
production of ammonia.

\subsection*{Comets}

To date, the only deuterated molecules observed in comets are HDO and
DCN\@. HDO was detected in Halley, Hyakutake, and Hale-Bopp, and the
observed HDO/H$_2$O ratios lie in the range 5.7--6.6$\times10^{-4}$
(e.g.\ Meier \etal 1998a). DCN was seen only in Hale-Bopp, where a
DCN/HCN ratio of 0.002 was determined (Meier \etal 1998b), although
Blake \etal (1998) showed that the DCN/HCN ratio was a factor of ten
larger in the dust jets. We have used our comet model to follow the
deuterium chemistry which occurs in the coma  and find that, as in
hot cores, post-evaporation chemistry does not significantly alter the
initial D/H ratios (Rodgers \& Charnley 2001d). Therefore, for parent
species we can use the observed coma ratios to infer the nuclear D/H
ratio, and for daughter species we can use these ratios to constrain the
nature of the parent and the chemical mechanism by which the species
are formed. In particular, if the DNC/HNC ratio can be determined in a
comet, we may be able to ascertain the origin of cometary HNC, which
is currently not well understood (see above).

Comparison of cometary D/H ratios with interstellar values may be used
to determine the physical conditions in the protosolar nebula, since
it is likely that the original interstellar values were altered by
chemical processing and/or mixing in the nebula (Aikawa \& Herbst
1999; Mousis \etal 2000). The observed D/H ratios in comets are
similar to the values in hot cores, but since HCN could possibly be a
daughter species in hot cores, this leaves only one molecule, HDO, for
which comparisons are currently possible. Future observations of other
deuterated cometary molecules are therefore vital. An exciting recent
development is the first tentative detection of HDCO in a comet,
Hale-Bopp (Kuan \etal 2001a). A determination of the HDCO/H$_2$CO
ratio, when combined with the known D/H ratios in water and hydrogen
cyanide, will further constrain models of cometary origin. It is also
useful to compare cometary D-enrichment with that seen in the Earth's
oceans -- it appears that comets have twice as much deuterium as the
Earth, which casts doubt on the assertion that comets brought the bulk
of the volatiles to the Earth. On the other hand, Delsemme (2000) has
argued that the Earth's water was deposited mainly by deuterium-poor
comets formed in the inner region of the protosolar nebula; such
comets should account for $\sim4$\% of the Oort cloud's population.

\subsection*{CARBON AND NITROGEN FRACTIONATION}

\subsection*{Interstellar Medium}

Heavy isotopes of carbon and nitrogen, $^{13}$C and $^{15}$N, are
formed in low- to intermediate-mass stars, and so their abundances
relative to $^{12}$C and $^{14}$N increase toward the galactic center
(e.g.\ Boogert \etal 2000). Langer \etal (1984) modelled the
fractionation in $^{13}$C occurring in interstellar molecules, and
concluded that it is preferentially locked up in CO, and hence
depleted in species such as C$^+$, H$_2$CO, CS, CH, and HCN\@.
Terzieva \& Herbst (2000) modelled the fractionation occurring in
$^{15}$N, and concluded that in typical dark clouds the resulting
fractionation is negligible. However, there is growing evidence for
the presence of `depletion cores' in some clouds, where most of the
molecules have frozen out onto grains, but some N$_2$ remains in the
gas phase (e.g.\ Charnley 1997c; Caselli \& Walmsley 1999; Bergin
\etal 2001). In this case, only simple N-bearing molecules are present
in the gas, and significant fractionation can occur (Charnley \&
Rodgers 2001).

\subsection*{Comets}

No significant fractionation is seen in cometary dust (Jessberger
1999), except for some grains which are strongly depleted in $^{13}$C,
which are thought to be unprocessed presolar grains formed in the
outflows of massive stars. In cometary volatiles, the
$^{12}$C/$^{13}$C ratios have been determined from observations of
C$_2$, CN, and HCN in a number of comets (Wyckoff \etal 2000), and are
found to lie in the range 85--93, consistent with the terrestrial
value of 90. The cometary $^{14}$N/$^{15}$N ratio has been measured in
CN and HCN (Crovisier \& \BoMo\ 1999), and is found to equal 323,
slightly larger than the terrestrial value of 272. However, the
$^{15}$N/$^{14}$N ratio is seen to vary amongst different solar system
objects, and the initial protosolar value is uncertain (e.g.\
Kallenbach \etal 1998). Large $^{15}$N enhancements are seen in some
meteoritic and interplanetary dust particle material (Messenger \&
Walker 1997), and the measurement of a large $\rm ^{15} NH_3/
^{14}NH_3 $ ratio in a comet would provide positive evidence for the
interstellar fractionation scenario proposed by Charnley \& Rodgers
(2001).

\section*{SULPHURETTED MOLECULES}

\subsection*{Interstellar Medium}

The sulphur chemistry in cold clouds is not fully understood. Gas
phase sulphur appears to be depleted in dark clouds by a factor of
100-1000 relative to its cosmic abundance, whereas it is undepleted in
the diffuse ISM (e.g.\ Ruffle \etal 1999). It is clear that somehow
sulphur must become preferentially incorporated into refractory
material as compared with C, N and O\@. The only solid phase
sulphuretted molecule detected to date is carbonyl sulphide (OCS)
which has an abundance relative to H$_2$O of 0.04--0.1\% (Palumbo
\etal 1997). In hot cores, a wide variety of S-bearing molecules are
observed.  Charnley (1997b) modelled the sulphur chemistry occurring
after the evaporation of H$_2$S, and showed that all the observed
sulphuretted species except OCS can be formed in the hot gas from this
one parent molecule. The predicted abundances of different S-bearing
daughter molecules depend on the core age and temperature, which means
that observations of species such as SO and SO$_2$ may be used as
`chemical clocks' to determine the ages of the cores (e.g.\ Hatchell
\etal 1998).

\subsection*{Comets}

To date, eight S-bearing molecules have been detected in comets, four
of them solely in Hale-Bopp (Crovisier \& \BoMo\ 1999). Many of these
are also observed in hot cores, so it is possible that sulphuretted
cometary molecules are formed in the coma from reactions of
H$_2$S\@. We have previously discussed the fact that the organic
molecules seen in Hale-Bopp cannot have been formed via coma
chemistry, which appears to suggest that the S-bearing molecules in
Hale-Bopp must also have been present in the nucleus. However, hot
core organic chemistry is driven by ion-molecule reactions, whereas
the sulphur chemistry is initiated by the reaction of H$_2$S and H to
form SH, which has a barrier of $\sim 860$\,K\@. The large abundances
of suprathermal H atoms in the inner coma may therefore be
instrumental in rapidly converting H$_2$S into other molecules (cf.\
Rodgers \& Charnley 1998).


\section*{CONCLUSIONS}

In this article we have presented a review of our current
understanding of the organic chemistry that can occur in dense
molecular clouds and in comets. By comparing the respective molecular
inventories of interstellar and cometary environments we can begin to
elucidate the degree of alteration of the interstellar material as it
becomes incorporated into protostellar nebulae.  In the next decade,
astronomical observations of distant star-forming regions, combined
with analyses of meteoritic and cometary samples, should allow the
relative degree of pristinity of cometary material to be determined.
Space-based telescopes of relevance to interstellar and cometary
chemistry include {\it SIRTF, FIRST-Herschel, NGST}, and {\it
SOFIA}\@. Although cometary H$_2$O can be detected from the ground
through radiatively pumped `hot-band' lines (Dello Russo \etal 2000),
space-based observatories are essential to overcome the telluric
extinction affecting most water lines. {\it SWAS} has observed H$_2$O
in comets Lee and McNaught-Hartley (Neufeld \etal 2000), and {\it
FIRST-Herschel} should allow the determination of HDO/H$_2$O in a
large sample of comets, including both short and long period comets.
The {\it Stardust} mission, although still some years from its
encounter with comet P/Wild~2, has already detected interstellar dust
particles.  Preliminary results indicate that these particles are very
similar to the complex organics seen in comet P/Halley, providing
further evidence for a link between interstellar and cometary
materials; such particles are understood to be destroyed in cometary
comae to provide the observed `extended sources' of molecules (e.g.\
Huebner 1987; Meier \etal 1993). Finally, the {\it Rosetta} mission
will be launched in 2003 and is due to rendezvous with comet
P/Wirtanen in 2011. The {\it Rosetta} lander module will allow the
first {\it in situ} examination of the nature of a cometary nucleus,
including the determination of the elemental, molecular and isotopic
composition of the surface and subsurface layers.  Assimilation and
interpretation of the data returned by each of these missions will
represent a significant advance in the quest for a detailed
understanding of the connection between interstellar and cometary
organic material.

\section*{ACKNOWLEDGMENTS}

Theoretical astrochemistry at NASA Ames is supported by NASA's Origins
of Solar Systems and Exobiology Programs through NASA Ames Interchange
NCC2-1162. SDR is supported by a National Research Council
postdoctoral research associateship.  The research of YJK was
supported by grants NSC 88-2112-M-003-013 and NSC 89-2112-M-003-004.
NRAO is operated by the Associated Universities, Inc., in cooperative
agreement with the National Science Foundation.


\end{document}